# On the digital holographic interferometry of fibrous material, I: Optical properties of polymer and optical fibers

K. M. Yassien[1], M. Agour[2,*], C. von Kopylow[2] and H. M. El Dessouky[3,4]

[1]Department of Physics, Aswan Faculty of Science, South Valley University, 81528 Aswan, Egypt

[2]Department of Optical Metrology, Bremer Institut für angewandte Strahltechnik, Klagenfurter Str. 2, D-28359 Bremen, Germany

[3]Fibers Research Lab. School of Design, University of Leeds, Leeds, LS2 9JT, UK

[4]Department of Physics, Faculty of Science, Mansoura University, Mansoura, Egypt

[*]Corresponding author: agour@bias.de

## Abstract

The digital holographic interferometry (DHI) was utilized for investigating the optical properties of polymer and optical fibers. The samples investigated here were polyvinylidene fluoride (PVDF) polymer fiber and graded-index (GRIN) optical fiber. The phase shifting Mach-Zehnder interferometer was used to obtain five phase-shifted holograms, in which the phase difference between two successive holograms is π/2, for each fiber sample. These holograms were recorded using a CCD camera and were combined to gain a complex wavefield, which was numerically reconstructed using the convolution approach into amplitude and phase distributions. The reconstructed phase distribution was used to determine the refractive index, birefringence and refractive index profile of the studied samples. The mean refractive index has been measured with accuracy up to $4 \times 10^{-4}$. The main advantage of DHI is to overcome the manual focusing limitations by means of the numerical focusing. The results showed accurate measurements of the optical properties of fibers.

**References**

1. N. Barakat and A. A. Hamza, "Interferometry of Fibrous Materials" (Adam Hilger, Bristol, 1990).
2. S. C. Simmens, "Birefringence Determination in Objects of Irregular Cross-sectional Shape and Constant Weight per Unit Length," Nature **181,** 1260-1261 (1958).
3. M. Pluta, "Interference microscope of polymer fibers," J. Microsc. **96**, 309–332 (1972).

4. A. A. Hamza, "A contribution to the study of optical properties of fibers with irregular transverse sections," Text Res. J. **50**, 731-734 (1980).

5. A. A. Hamza, T. Z. N. Sokkar, Ghander A. M., M. A.Mabrouk and W. A. Ramadan, "On the determination of the refractive index of a fibre. II. Graded index fibre," Pure. Appl. Opt. **4**, 161-177 (1995).

6. A. A. Hamza, A. E. Belal, T. Z. N. Sokkar, M. A. El-Bakary and K. M. Yassien, "Measurement of the spectral dispersion curves of low birefringence polymer fibres," Opt. Laser Eng. **45,** 922–928 (2007).

7. Z. Liu, X. Dong, Q. Chen, C. Yin, Y. Xu, and Y. Zheng, "Nondestructive measurement of an optical fiber refractive-index profile by a transmitted-light differential interference contact microscope," Appl. Opt.**43,** 1485-1492 (2004).

8. K. M. Yassien, "Comparative study on determining the refractive index profile of polypropylene fibres using fast Fourier transform and phase-shifting interferometry," J. Opt. A: Pure Appl. Opt. **11**, 075701(2009).

9. U. Schnars and W. Jüptner, Digital Holography (Springer-Verlag, Berlin, 2005).

10. D. Gabor, "A new microscopic principle," Nature **161**, 777–778 (1948).

11. K. A. Stetson and R. L. Powell, "Interferometric hologram evaluation and real-time vibration analysis of diffuse objects," J. Opt. Soc. Am. **55**, 1694-1695 (1965).

12. U. Schnars and W. Jüptner, "Digital recording and numerical reconstruction of holograms," Meas. Sci. Technol. **13**, R85-R101 (2002).

13. L. P. Yaroslavsky, Digital Holography and Digital Image Processing, Principles, Methods, Algorithms (Kluwer Academic Publishers, Boston, 2004).

14. U. Schnars and W. Jüptner, "Direct recording of holograms by a CCD target and numerical reconstruction," Appl. Opt. **33**, 179-181 (1994).

15. M. Hossain, D. S. Mehta, C. Shakher , "Refractive index determination: an application of lensless Fourier digital holography," Optical Eng. **45,** 106203 (2006).

16. M. de Angelis, S. De Nicola, A. Finizio, and G. Pierattini, "Digital-holography refractive-index-profile measurement of phase gratings," Appl. Phys. Lett. **88**, 111114 (2006).

17. H.. Wahba and T. Kreis, "Characterization of graded index optical fibers by digital holographic interferometry, " Appl. Opt. **48**, 1573-1582 (2009)

18. T. Kreis, "Holographic Interferometry" (Akademie Verlage, Berlin, 1996).

19. M. A. El-Morsy, T. Yatagai, A. A. Hamza, M. A. Mabrouk, T. Z. N. Sokkar, "Automatic refractive index profiling of fibers by phase analysis method using Fourier transform," Opt. Laser Eng. **38**, 509–525 (2002).

20. M. Pluta, "Advanced Light Microscopy," (PWN, Warszawa, 1993).

21. C. Koliopoulos "Interferometric Optical Phase Measurement Techniques," Ph.D. Dissertation (U. Arizona: Optical Sciences Center, 1981).

22. Y. Chengand, J. Wyant, "Phase shifter calibration in phase-shifting interferometry," App. Opt. **24**, 3049-3052 (1985).

23. H. Zanger, C. Zanger, "Fibre optics communication and other application," (New York: Maxwell Macmillan International Publishing, 1991).

24. R. Benjamin, M. Pierre, C. Etienne, E. Yves, D. Christian, J. M. Pierre, "Measurement of the integral refractive index and dynamic cell morphometry of living cells with digital holographic microscopy," Optics Express **13**, 9361-9373 (2005).

25. Y. Surrel, "Design of algorithms for phase measurements by the use of phase stepping," Appl. Opt. **35,** 51-60 (1996).

26. H. Zhang, M. Lalor, D. Burton, "Error-compensating algorithms in phase-shifting interferometry: a comparison by error analysis," Opt. Laser Eng. **31**, 381-400 (1999).

27. K. Hibino and M. Yamauch, "Phase-measuring algorithms to suppress spatially nonuniform phase modulation in a two-beam interferometer," Optical Review **7**, 543-549 (2000).

28. H. M. El-Dessouky , C. A. Lawrence, A. M. Voice, E. L. V. Lewis and I. M. Ward, "An Interferometric Prediction of the Intrinsic Optical Properties for Cold-drawn iPP, PTFE and PVDF Fibres"  J. Opt. A: Pure & Appl. Opt. **9**, 1041-1047 (2007).

29. A. A. Hamza, A. E. Belal, T. Z. N. Sokkar, H. M. El-Dessouky and M. A. Agour, "Interferometric Studies on the Influence of Temperature on the Optical and Dispersion Parameters of GRIN Optical Fibres", Optics and Laser in Eng, **45**, 145-152 (2007).